\documentclass[onecolumn,notitlepage,superscriptaddress,preprintnumbers,amsmath,aps,pra,11pt]{revtex4-1}
\usepackage{graphicx}
\usepackage{color}
\usepackage{subfigure}
\usepackage[usenames,dvipsnames]{xcolor}

\usepackage{hyperref}

\begin{document}
%\preprint{\large{Draft: \today}}
\title{Distinguishing between non-orthogonal quantum states of a single spin}

\affiliation{3. Physikalisches Institut, Research Center SCOPE, and MPI for Solid State Research, University of Stuttgart, Pfaffenwaldring 57, 70569 Stuttgart, Germany}
\affiliation{SUPA, School of Engineering and Physical Sciences, Heriot-Watt University, Edinburgh EH14 4AS, United Kingdom}
\affiliation{Institut f\"{u}r Quantenoptik, Universit\"{a}t Ulm, 89073 Ulm, Germany}

\author{Gerald Waldherr}
\affiliation{3. Physikalisches Institut, Research Center SCOPE, and MPI for Solid State Research, University of Stuttgart, Pfaffenwaldring 57, 70569 Stuttgart, Germany}

\author{Adetunmise C. Dada}
\affiliation{SUPA, School of Engineering and Physical Sciences, Heriot-Watt University, Edinburgh EH14 4AS, United Kingdom}

\author{Philipp Neumann}
\affiliation{3. Physikalisches Institut, Research Center SCOPE, and MPI for Solid State Research, University of Stuttgart, Pfaffenwaldring 57, 70569 Stuttgart, Germany}

\author{Fedor Jelezko}
\affiliation{Institut f\"{u}r Quantenoptik, Universit\"{a}t Ulm, 89073 Ulm, Germany}

\author{Erika Andersson}
\affiliation{SUPA, School of Engineering and Physical Sciences, Heriot-Watt University, Edinburgh EH14 4AS, United Kingdom}

\author{J\"org Wrachtrup}
\affiliation{3. Physikalisches Institut, Research Center SCOPE, and MPI for Solid State Research, University of Stuttgart, Pfaffenwaldring 57, 70569 Stuttgart, Germany}

\begin{abstract}
An important task for quantum information processing is optimal discrimination between two non-orthogonal quantum states, which until now has only been realized optically. 
Here, we present and compare experimental realizations of optimal quantum measurements for distinguishing between two non-orthogonal quantum states encoded in a single $^{14}$N nuclear spin. 
Implemented measurement schemes are the minimum-error measurement (known as Helstrom measurement), 
unambiguous state discrimination using a standard projective measurement, and optimal unambiguous state discrimination (known as IDP measurement), which utilizes a three-dimensional Hilbert space.
Measurement efficiencies are found to be above 80\% for all schemes and reach a value of 90\% for the IDP measurement. 
\end{abstract}

\maketitle
          
%\section{Introduction}
Due to imperfections, real measurements on quantum systems are seldom ideal projective measurements. 
Moreover, it is sometimes advantageous to deliberately design a quantum measurement that is not projective, depending on exactly what property of the measured quantum system one is interested in.
 One example is when distinguishing between non-orthogonal quantum states. This is not possible to achieve perfectly with certainty. Nevertheless, we can for example minimize the error~\cite{helstrom}, 
or measure in such a way that when a result is obtained it is guaranteed to be correct. This can be achieved at the expense of sometimes obtaining an inconclusive result, and such a measurement is referred to as unambiguous~\cite{bergoubk}.

 An optimal quantum measurement is  often not a standard projective quantum measurement,  but a so-called generalized quantum measurement. While the minimum-error measurement when distinguishing between two equiprobable nonorthogonal states is a projective measurement known as the Helstrom measurement, 
the optimal unambiguous state discrimination (USD) measurement is a generalized measurement known as the Ivanovic-Dieks-Peres (IDP) measurement~\cite{Ivanovic1987257,Dieks1988303,Peres198819}.

This type of measurement is relevant for important quantum information tasks in quantum cryptography~\cite{PhysRevA.66.042313,PhysRevLett.68.3121} and in entanglement swapping protocols~\cite{PhysRevA.71.012303}.
Moreover, it can be useful for quantum communication, when the two signal states are non-orthogonal after passing through a channel.
Yet, while it has been implemented optically~\cite{PhysRevA.63.040305,PhysRevA.83.042339}, an implementation in solid state has been lacking until now. In order to perform advanced QIP tasks, the ability to perform such optimal (generalized) 
quantum measurements is a basic requirement.

Our experiments are carried out on the $^{14}$N nuclear spin (spin $I=1$) of a negatively charged nitrogen-vacancy (NV$^-$) defect in diamond, utilizing a quantum non-demolition (QND) single shot readout method \cite{PhilippNeumann07302010}. 
The NV$^-$ defect consists of a substitutional nitrogen atom and a neighbouring vacancy site inside the diamond lattice, and has many favorable features, which make it a promising system for quantum information processing (QIP) \cite{Gruber_Science1997, Jelezko_PRL2004, Childress_Science2006, Balasubramanian, PhilippNeumann07302010, Robledo_Nature2011}.

\begin{figure}
\centerline{\includegraphics[width=0.45\textwidth]{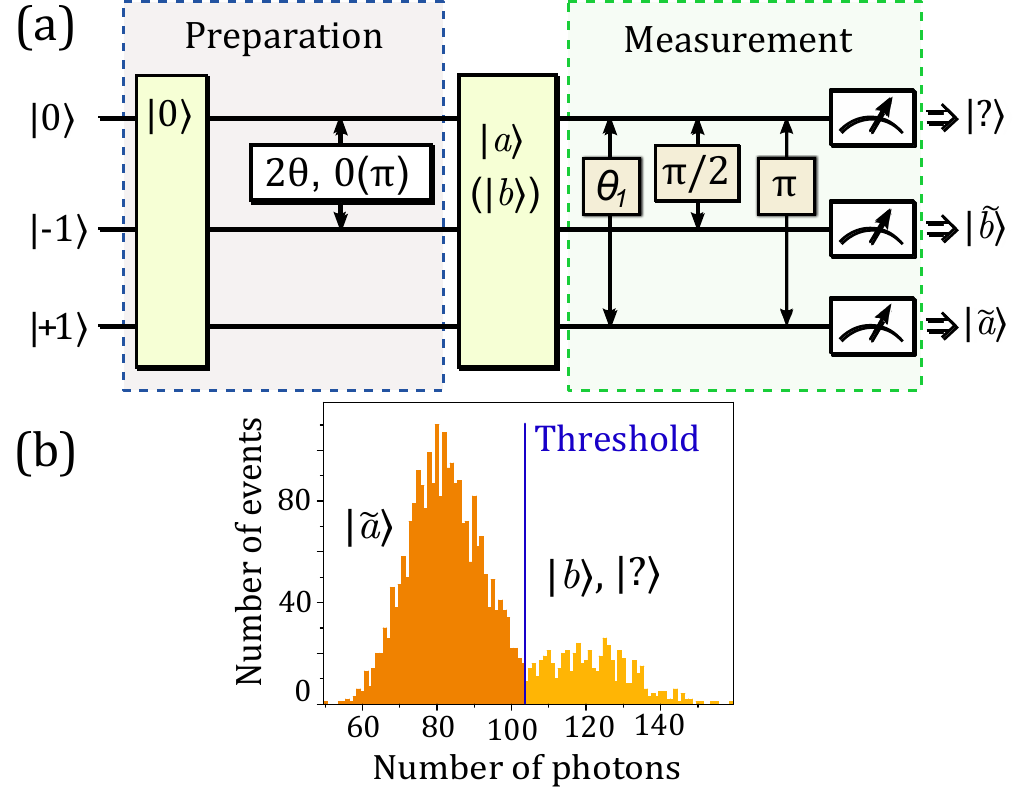}}
\caption{(a) Illustration of the applied radio frequency pulses to prepare the non-orthogonal states and perform the IDP measurement. For the Helstrom measurement, we omit the $\theta_1$ pulse. (b) Histogram of results for single shot readout of $|\tilde{a}\rangle$ for several preparations of a single NV center in the state $|a\rangle$ with $\theta=\pi/4$. In each event (preparation and detection), a level of photocounts below the threshold is taken to correspond to the system having been prepared in the state $|a\rangle$.}
\label{fig:nvfig1}
\end{figure}
For both measurements, we consider the case where the system has been prepared, with equal prior probabilities $p_a=p_b=1/2$, in one of the two non-orthogonal states
\begin{equation}
\label{eq:idp1}
|a\rangle = \cos\theta | 0\rangle - \sin\theta |{\rm-}1 \rangle,~~
|b\rangle = \cos\theta| 0\rangle + \sin\theta |{\rm-}1 \rangle,
\end{equation}
where $0\le\theta\le \pi/4$, and $\theta$ is half the angle between $|a\rangle$ and $|b\rangle$, and ${|0\rangle, |1\rangle}$ form an orthonormal basis.
Since the states $|a\rangle$ and $|b\rangle$ are not orthogonal, they cannot be distinguished from each other with certainty. 

We prepare these states experimentally on the $^{14}$N nuclear spin.
First, we initialize to the $|m_{\rm{I}}=0\rangle$ state by QND measurement \cite{PhilippNeumann07302010}. This is successful whenever the measurement result is $m_I=0$. 
Additionally, we perform charge state postselection on the NV as described in \cite{Waldherr_PRL2011b}, and the state is successfully prepared only if the NV is found in its negative charge state.
Next we prepare the two non-orthogonal states~\eqref{eq:idp1} by applying a $2\theta$-pulse resonant with the $|m_{\rm{I}}=0\rangle$ $\leftrightarrow$  $|m_{\rm{I}}=-1\rangle$ transition, with phases $0$ and $\pi$, producing states $|a\rangle$ and $|b\rangle$ respectively, see Fig.~\ref{fig:nvfig1}.
We then try to distinguish between these two states with different measurement protocols.
Essentially, the protocol tells us which measurement basis to use.
To implement this in the experiment, we apply radio frequency (RF) pulses resonant with the $^{14}N$ spin transitions $|m_{\rm{I}}=0\rangle$ $\leftrightarrow$  $|m_{\rm{I}}=-1\rangle$ and $|m_{\rm{I}}=0\rangle$ $\leftrightarrow$  $|m_{\rm{I}}=+1\rangle$, such that the required basis is rotated onto the $\{|m_{\rm{I}}=0\rangle, |m_{\rm{I}}=+1\rangle, |m_{\rm{I}}=-1\rangle\}$-basis.
 The final projective measurements are performed by consecutive single-shot readout measurements on the spin states, where the first positive result is counted as the outcome. Due to the possibilities of spin flips and/or making errors in each readout measurement in the sequence, there is a probability to obtain multiple or no positive result at all leading to imperfect detection efficiencies. Details for specific cases are given below.

%\section{Unambiguous state discrimination}
In USD, we require $p(a|b)=p(b|a)=0$, where $p(a|b)$ is the probability to obtain result ``$a$" given that the state was $|b\rangle$, and vice versa for $p(b|a)$.
Trivially, this can be achieved by a projective measurement in the two-dimensional space spanned by $|0\rangle$ and $|-1\rangle$, either in the basis $\{|a\rangle, |a^\perp\rangle\}$ or the basis $\{|b\rangle, |b^\perp\rangle\}$, where $|a^\perp\rangle=\sin\theta|0\rangle+\cos\theta |{\rm-}1\rangle$ is orthogonal to $|a\rangle$, and $|b^\perp\rangle=\sin\theta |0\rangle - \cos\theta |{\rm-}1\rangle$ is orthogonal to $|b\rangle$. 
The result ``$a^\perp$" then guarantees that the state must have been $|b\rangle$, and vice versa for ``$b^\perp$", while the results ``$a$" and ``$b$" are inconclusive. 
In the following, we call this method standard unambiguous state discrimination (SUSD).
The probability of such an inconclusive outcome is $p_?=p_i+p_j|\langle i|j\rangle|^2$, $i, j = a, b$.
If the prior probabilities $p_a$ and $p_b$ are equal, then $p_?=(1+|\langle a|b\rangle|^2)/2\ge 1/2$ for either measurement.
If the prior probabilities are not equal, it will be best to always choose the one with the higher probability.
The main drawback of this protocol is the high probability of an inconclusive result.

To realize the SUSD measurement in the basis $\{|a\rangle, |a^\perp\rangle\}$ and in the basis $\{|b\rangle, |b^\perp\rangle\}$, we first perform unitary operations $\hat U_{a}$ and $\hat U_{b}$ respectively, and then a detection in the $\{|0\rangle, |1\rangle\}$ basis. 
The unitary operations $\hat U_{a}=|0\rangle\langle a|+|1\rangle\langle a^\perp|$ and $\hat U_{b}=|0\rangle\langle b|+|1\rangle\langle b^\perp|$ are carried out by RF $2\theta$ and $-2\theta$-pulses, respectively, on the $|m_{\rm{I}}=0\rangle \leftrightarrow |m_{\rm{I}}=-1\rangle $ transition.
Additionally, we apply a $\pi$-pulse on the $|m_{\rm{I}}=0\rangle \leftrightarrow |m_{\rm{I}}=+1\rangle $ transition (see error discussion), and the final projective measurement is performed in the basis \{$|m_{\rm{I}}=+1\rangle, |m_{\rm{I}}=-1\rangle$\}. 
For this measurement, the probability to obtain a result was $\sim$84.6 \% (analogous to the detection efficiency with photons). This includes $\sim$1\% of multiple positive results.
The state $|m_{\rm{I}}=-1\rangle$ then corresponds to $|a^\perp\rangle$ and $|b^\perp\rangle$, depending on the chosen basis, and the state $|m_{\rm{I}}=+1\rangle$ corresponds respectively to $|a\rangle$ and $|b\rangle$, the inconclusive result.

\begin{figure}
\centerline{\includegraphics[width=0.5\textwidth]{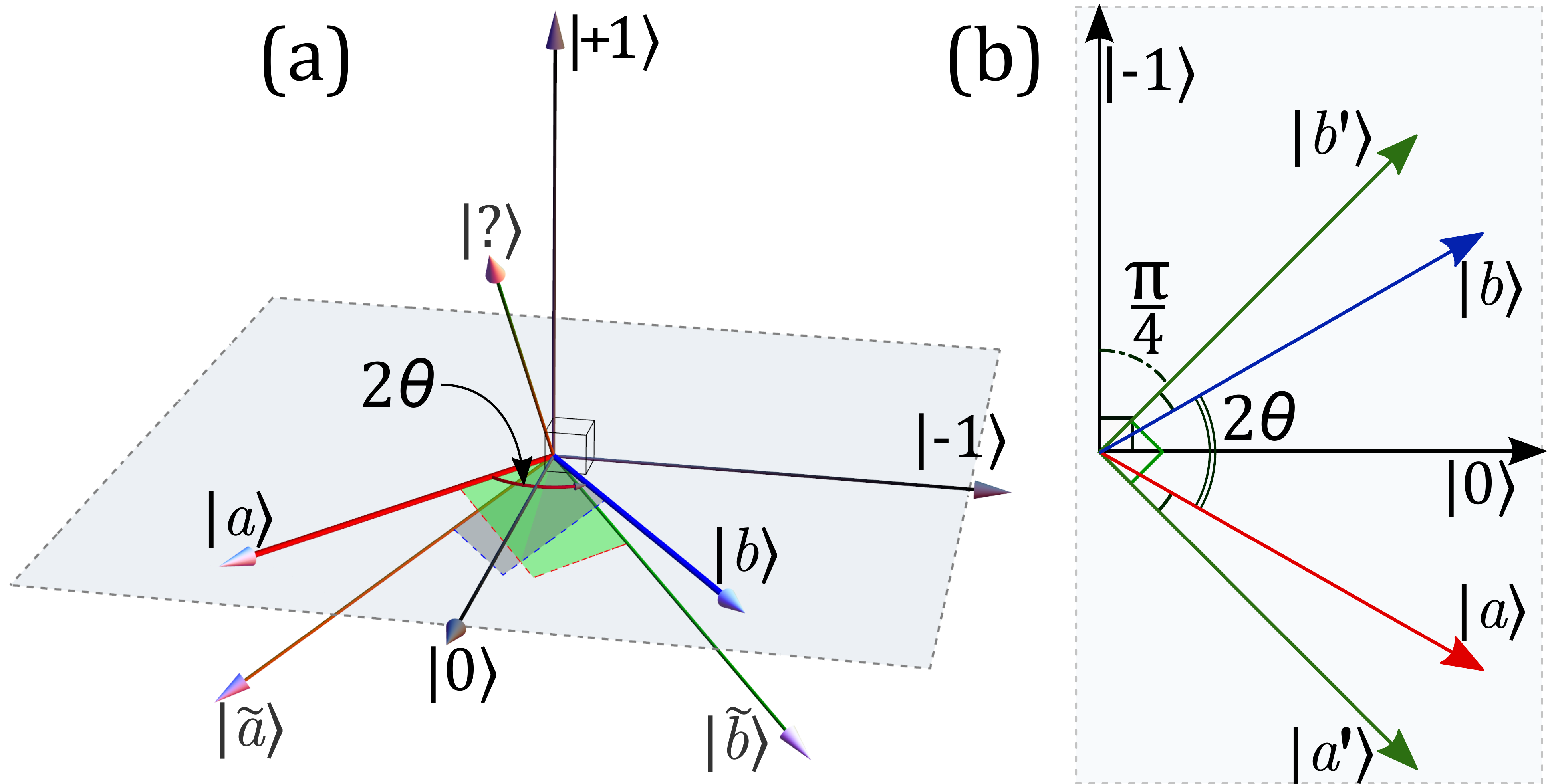}}
\caption{ Geometrical representation of (a) the IDP measurement. The initial two-dimensional Hilbert space, where $|a\rangle$ and $|b\rangle$ live, is spanned by \{$|0\rangle$, $|{\rm-}1\rangle$\}.  The measurement is a projective measurement in the basis $\{|\tilde{a}\rangle |\tilde{b}\rangle, |?\rangle\}$, where $|\tilde{a}\rangle$ is orthogonal to $|b\rangle$ and $|\tilde{b}\rangle$ is orthogonal to $|a\rangle$. It can be realized by first making the unitary operation $\hat{U}$ in Eq.~\eqref{eq:uagain3}, followed by a projective measurement in the \{$|0\rangle$, $|{\rm-}1\rangle$, $|{\rm+}1\rangle$\} basis. (b) The Helstrom measurement basis $\{|a'\rangle,|b'\rangle\}$.}
\label{fig:geomreps}
\end{figure}

The optimal USD (IDP) however has three outcomes, corresponding to ``$a$", ``$b$", and ``inconclusive".
It is optimal in the sense that the probability of an inconclusive result is the lowest possible for unambiguous results of ``$a$" and ``$b$".
The measurement can be understood as a projective measurement in an extended three-dimensional space, as shown in Fig.~\ref{fig:geomreps}(a). 
This requires an auxiliary basis state which is provided by the $^{14}$N nuclear spin, which has three spin states ($I$=1). In three dimensions, as Fig.~\ref{fig:geomreps} (a) shows, we can find orthonormal states $|\tilde{a}\rangle, |\tilde{b}\rangle$ and $|?\rangle$, so that $|\tilde{a}\rangle$ is perpendicular to $|b\rangle$, and $|\tilde{b}\rangle$ is perpendicular to $|a\rangle$. A possible choice of such states is given by
\begin{eqnarray}
|\tilde{a}\rangle&=&\frac{1}{\sqrt 2}\left(\tan \theta |0\rangle-|{\rm-}1\rangle -\sqrt{1-\tan^2\theta}|{\rm+}1\rangle \right),\nonumber\\
|\tilde{b}\rangle&=&\frac{1}{\sqrt 2}\left(\tan \theta |0\rangle+|{\rm-}1\rangle -\sqrt{1-\tan^2\theta}|{\rm+}1\rangle \right),\nonumber\\
|?\rangle&=& \sqrt{1-\tan^2\theta}|0\rangle+\tan\theta|{\rm+}1\rangle .
\end{eqnarray}
For a projective measurement in the basis $\{|\tilde{a}\rangle, |\tilde{b}\rangle, |?\rangle\}$, it holds that $p(a|b)=p(b|a)=0$ as required, and the probability $p_?$ for an inconclusive outcome is given by the overlap $|\langle a| b\rangle|$. 
The Hilbert space can be extended either using an ancillary level, as done here, or using an ancillary qubit, as done in optical realizations. 
A general method for working out how to realize generalized quantum measurements is given in Ref.~\cite{PhysRevA.63.052301}. The measurement in the basis $\{|\tilde{a}\rangle |\tilde{b}\rangle, |?\rangle\}$ can be implemented for example by first performing a unitary operation $~\hat U = |0\rangle \langle\tilde{a}|+|{\rm-}1\rangle \langle\tilde{b}|+|{\rm+}1\rangle \langle ?|,~$ 
followed by a projective measurement in the \{$|0\rangle$, $|{\rm{-}}1\rangle$, $|\rm{+}1\rangle$\} basis. 
Detection in $|0\rangle$ or  $|{\rm-}1\rangle$ now unambiguously indicates that the unknown state was $|a\rangle$ or $|b\rangle$ respectively, while the  result $|{\rm+}1\rangle$ is inconclusive. The conditional probabilities for different results are
\begin{eqnarray}
\label{eq:poutIDP}
	p(a|a)&=&|\langle \tilde{a}|a\rangle|^2=1-|\langle a|b\rangle|=p(b|b)=|\langle \tilde{b}|b\rangle|^2 \nonumber \\
	p(a|b)&=&p(b|a) =|\langle \tilde{a}|b\rangle|^2=|\langle \tilde{b}|a\rangle|^2=0  \nonumber \\
         p(?|a)&=&p(?|b)=  |\langle a|b\rangle|.
\end{eqnarray}
Here $p(j|k)$ denotes the conditional probability of obtaining a result $|j\rangle$ given a state $|k\rangle$, with $j,k=a,b$. The average probability of correctly identifying a prepared state is then $p_{\rm corr}=p(a|a)p_a+p(b|b)p_b$, where $p_j$ are the prior probabilities, i.e., for preparing states $|j\rangle$. Also, the average error probability is $p_{\rm err}=p(b|a)p_a+p(a|b)p_b$, while the average probability of an inconclusive outcome is $p_?=p(?|a)p_a+p(?|b)p_b$.

Defining the basis vectors as $|0\rangle\equiv[1,0,0]^T$, $|{\rm-}1\rangle\equiv[0,1,0]^T$ and $|+1\rangle\equiv[0,0,1]^T$, where the superscript $T$ denotes transpose, $\hat U$ is given by 
\begin{equation}
\label{eq:uagain3}
 \hat{U} = \frac{1}{\sqrt 2}\left[ {\begin{array}{*{20}c}
    	\tan\theta & -1 & -\sqrt{1-\tan^2\theta} \\
   	\tan\theta & 1& -\sqrt{1-\tan^2\theta} \\
	\sqrt{2(1-\tan^2 \theta)} & 0 & \sqrt 2\tan\theta \\
\end{array}} \right]. 
\end{equation}

This $\hat{U}$ may be decomposed into a product of unitary operators coupling two levels at a time~\cite{PhysRevLett.73.58, PhysRevA.63.052301, PhysRevA.83.042339}. Here we choose, for ease of experimental realization, a decomposition of the form
$\hat{U}=\hat{T}_{0,-1}\hat{T}_{0,+1}$, 
where
\begin{align}
\hat{T}_{0,-1} = &\frac{1}{\sqrt 2}\left( \begin{array}{ccc}
1 & -1& 0 \\
1 & 1 & 0 \\
0 & 0 & \sqrt 2 \end{array} \right),\label{eq:theta2}\\
\hat{T}_{0,+1} =&\left[ {\begin{array}{*{20}c}
	\tan\theta   & 0 & -\sqrt{1-\tan^2\theta}\\
   	0 & 1 & 0  \\
	\sqrt{1-\tan^2\theta}& 0 & \tan\theta\\
\end{array}} \right].\label{eq:theta1}
\end{align}
This corresponds to a pulse sequence consisting of a $\theta_1=2 \arcsin (\sqrt{1-\tan^2\theta})$-RF pulse resonant with the $|m_{\rm{I}}=0\rangle$  $\leftrightarrow$ $|m_{\rm{I}}={\rm+}1\rangle $ transition, followed by a $\theta_2=\pi/2$ pulse resonant with the $|m_{\rm{I}}=0\rangle$  $\leftrightarrow$ $|m_{\rm{I}}={\rm-}1\rangle $ transition. 
As before, we also apply a RF $\pi$-pulse on the $|m_{\rm{I}}=0\rangle \leftrightarrow |m_{\rm{I}}=+1\rangle $ transition, such that if the RF pulses have no effect due to improper electron initialization, the measurement result will be inconclusive (the spin will stay in $|m_{\rm{I}}=0\rangle$, which gives an inconclusive result, see Fig. \ref{fig:nvfig1}).
The detection is completed with a projective measurement in the basis \{$|m_{\rm{I}}=0\rangle$,$|m_{\rm{I}}={\rm-}1\rangle$,$|m_{\rm{I}}={\rm+}1\rangle$\}.
A positive result on $|m_{\rm{I}}=-1\rangle$ ($|m_{\rm{I}}=+1\rangle$) corresponds to state $|a\rangle$ ($|b\rangle$), and a positive result on $|m_{\rm{I}}=0\rangle$ corresponds to an inconclusive result.
 In this case, the average detection efficiency of the final measurement is $\sim$90.2\%, including $\sim$10.2\% of multiple positive results.

\begin{table}[ht]
\centering    
\begin{tabular}{|c |c |c |c|c|}   
\hline
Method & $~~d~~$ &~unambiguous~&~~error \%~~&~~efficiency \%~~\\[0.5ex]  
\hline  \hline   
SUSD & 2 & yes & $\sim3.5$ & 84.6\\   \hline    
IDP & 3 & yes & $4-7.5$ & 90.2 \\\hline 
Helstrom & 2 & no & $> 3.5$ & 83.1\\
\hline     
\end{tabular}
\caption{Overview over the three measurements. Column $d$ is needed dimension of the Hilbert space. The Helstrom measurement has an inherent error probability; here we show the minimum error due to measurement imperfections. \label{tab:table1}} 
\end{table}

%\section{Minimum-error measurement}
The Helstrom measurement minimizes the error in the result in the case where inconclusive outcomes are not allowed. It gives
 a higher probability to obtain a correct result than an unambiguous measurement, but in return, an obtained result is not guaranteed to be correct. For the two equiprobable states $|a\rangle$ and $|b\rangle$ we are considering,
it is a projective measurement in a two-dimensional orthonormal  basis $\{|a'\rangle, |b'\rangle\}$ which is symmetric around $|a\rangle$ and $|b\rangle$ (see Fig. \ref{fig:geomreps} (b)), such that $|a\rangle$ has a larger overlap with $|a'\rangle$ and $|b\rangle$ a larger overlap with $|b'\rangle$.
In our case, $|a'\rangle=\frac{1}{\sqrt{2}}\left(|0\rangle-|-1\rangle\right)$,  $|b'\rangle=\frac{1}{\sqrt{2}}\left(|0\rangle+|-1\rangle\right)$. 
The state $|a'\rangle$ corresponds to $|a\rangle$, and $|b'\rangle$ to $|b\rangle$. 
For two states $|a\rangle$ and $|b\rangle$ with prior probabilities $p_a$ and $p_b$,  the states $|a'\rangle$ and $|b'\rangle$ are modified, and the Helstrom measurement has an error probability given by~\cite{helstrom}
\begin{equation}
\label{eq:helsbnd}
p_{\rm err}^{\rm opt}=\frac{1}{2}(1-\sqrt{1-4p_ap_b|\langle a| b\rangle|^2}).
\end{equation}
 The probability to obtain a correct result is given by $p_{\rm corr}=1-p_{\rm err}^{\rm opt}$, 
which is the highest possible probability to identify the state correctly.
For two equiprobable states $|a\rangle$ and $|b\rangle$ the Helstrom measurement is implemented by performing only the rotation \eqref{eq:theta2}, i.e, the RF $\pi/2$-pulse on the $|m_{\rm{I}}=0\rangle \leftrightarrow |m_{\rm{I}}=-1\rangle $ transition.
Again, we additionally apply the RF $\pi$-pulse on the $|m_{\rm{I}}=0\rangle \leftrightarrow |m_{\rm{I}}=+1\rangle $ transition. 
Finally, we do a projective measurement in the basis \{$|m_{\rm{I}}=+1\rangle$,$|m_{\rm{I}}={\rm-}1\rangle$\}, corresponding to the outcomes ``$a$" and ``$b$", 
with an efficiency of $\sim$83.1\%, including $\sim$1.1\% of multiple positive results.

\begin{figure*}[t]
\includegraphics[width=1.00\textwidth]{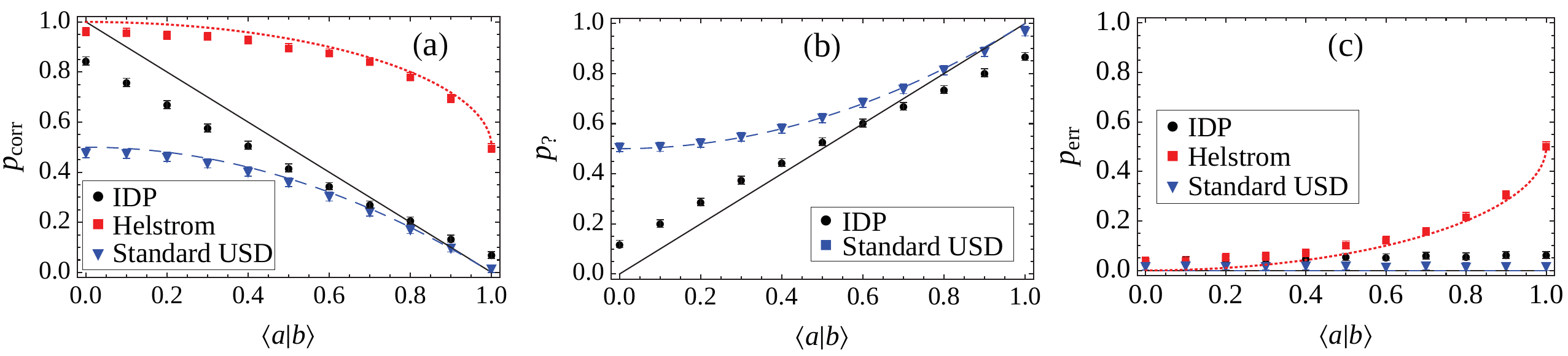}
	\caption{Experimental results of unambiguous and minimum-error discrimination of non-orthogonal states. The two non-orthogonal states $|a\rangle$ and $|b\rangle$ are prepared with equal prior probabilities, $p_a=p_b=1/2$. The probabilities for correct identification $p_{\rm corr}=p(a|a)p_a+p(b|b)p_b$, inconclusive outcomes $p_{\rm ?}=p(?|a)p_a+p(?|b)p_b$ and incorrect outcomes $p_{\rm err}=p(b|a)p_a+p(a|b)p_b$ are shown in figures(a), (b) and (c) respectively. Probabilities are plotted as functions of overlap $\langle a|b\rangle=\cos 2\theta$. The experimental error bars account only for uncertainty due to measurement shot noise.} 
\label{fig:resprobs}
\end{figure*}

%\section{Results}
An overview over the three measurement protocols in shown in Table~\ref{tab:table1}.
The results of the three measurements are shown in Fig.~\ref{fig:resprobs}, where we plot the average probability of correctly identifying a prepared state $p_{\rm corr}=p(a|a)p_a+p(b|b)p_b$, the  average probability for an inconclusive result $p_{\rm ?}=p(?|a)p_a+p(?|b)p_b$, and the average probability for conclusive but incorrect identification $p_{\rm err}=p(b|a)p_a+p(a|b)p_b$ (i.e. making an error).
The conditional probabilities $p(j|a)$ and $p(j|b)$ $(j\in\{a,b,?\})$ are found to be similar in our experiments.

First, we will compare the generalized quantum measurement (IDP) with the simpler SUSD measurement, to see how the increased experimental complexity affects the performance of the protocol.
In Fig.~\ref{fig:resprobs} (a) and (b) we see that the probability of a conclusive result and the probability for an inconclusive result is always better for the IDP measurement than for the SUSD measurement, as expected.
However, Fig.~\ref{fig:resprobs} (c) shows that this is also partly due to a higher error probability for the IDP measurement for large $\langle a| b\rangle$.
The reason for these errors is discussed below, and is expected to be reduced by future development in the experimental control of the NV.

Whereas the Helstrom measurement minimizes errors when no inconclusive results are allowed, USD measurements are supposed to give unambiguous (error-free) results by allowing an inconclusive outcome.  However, the results of the experimentally realized unambiguous measurements are not guaranteed to be correct either, due to inevitable experimental imperfections. It is therefore important to check how the errors resulting from imperfections in the realization of the unambiguous measurement compare with the error probability of the ideal optimal minimum-error measurement, as well as with the error probability in an experimental realization of the minimum-error measurement. 

As expected, we find in Fig.~\ref{fig:resprobs} (a) that the probability of obtaining a correct result is highest for the Helstrom measurement.
The probability of making an error is shown in Fig.~\ref{fig:resprobs} (c).
We can see that for small overlap $|\langle a| b\rangle|$, this probability is only a few percent for the three measurement protocols. Also, the error in the implementation of the IDP measurement is never greater than that of the Helstrom measurement. Only for small overlap does the error in the implementation of the IDP measurement slightly exceed the ideal minimum-error bound.

%\section{Error discussion}
The main source of errors, which lead to discrepancies between ideal theoretical bounds and the experimental results, is the limited lifetime of the nuclear spin during the QND readout.
On one hand, this limits the available time and therefore also the photon count for single shot readout, which increases errors due to photon shot noise.
For example, even though the spin is in the state $|m_{\rm{I}}={\rm+}1\rangle$, the count rate corresponding to state $|m_{\rm{I}}=0\rangle$ or $|m_{\rm{I}}={\rm-}1\rangle$ is detected. 
On the other hand, there is the possibility that the spin state flips during the measurement, which is especially important for consecutive measurements on all three spin states.
Due to these errors, we can get more or less than one positive result for the measurement. Therefore, we count the first positive result as the outcome.

 The probability to get a false positive result is increased if a neighboring spin state (with $\Delta m_{\rm{I}} = \pm 1$) is highly populated.
This is why the error probability of the IDP measurement is higher for large $\langle a|b\rangle$ compared to the standard unambiguous measurement in Fig.~\ref{fig:resprobs} (c), since for the IDP measurement the inconclusive result corresponds to $|m_{\rm{I}}=0\rangle$ and the conclusive result  $|m_{\rm{I}}={\rm-}1\rangle$ or $|m_{\rm{I}}={\rm+}1\rangle$, whereas for the standard unambiguous measurement the 
inconclusive result corresponds to $|m_{\rm{I}}={\rm+}1\rangle$ and the conclusive result $|m_{\rm{I}}={\rm-}1\rangle$. 

Another point to consider is the imperfect electronic initialization, either due to errors in the charge state postselection (cf. \cite{Waldherr_PRL2011b}), or because the electron spin is not properly polarized.
In both cases, this affects the nuclear spin transition frequencies by hyperfine interaction, and the RF pulses, which are designed for the electron being in the $m_{\rm{S}}=0$ NV$^-$ ground state, will have no effect on the nuclear spin.
In this case, the nuclear spin will stay in the initialized state, which is $m_{\rm{I}}=0$.
By applying a final $\pi$ pulse on the $|m_{\rm{I}}=0\rangle \leftrightarrow |m_{\rm{I}}=+1\rangle $ transition, we ensure that the state $|m_{\rm{I}}=0\rangle$ is either not used (for the standard unambiguous and Helstrom measurements), or counted as inconclusive (for the IDP measurement).

%\section{Conclusion}
In conclusion, we have experimentally realized and compared three different measurement protocols to distinguish between non-orthogonal quantum states, including optimal unambiguous state discrimination and the minimum-error or Helstrom measurement.
Previously, optimal unambiguous state discrimination has only been realized optically~\cite{PhysRevA.63.040305,PhysRevA.83.042339}. The ability to perform generalized measurements on NV$^-$ centers is of interest for implementations of solid-state quantum computing.  
The realized IDP measurement for NV$^-$  centers outperforms standard projective measurements, and gives further evidence that NV$^-$ centers in diamond are a favorable candidate for solid state quantum information processing  at room temperature.

We are grateful for financial support by EU project SOLID and SQUTEC and by the MPI Fellowship.
AD gratefully acknowledges funding from the Scottish Universities Physics Alliance (SUPA) and useful initial discussions with Johannes Beck and Florian Rempp. EA acknowledges partial support by EPSRC EP/G009821/1.

%\bibliography{allreferences1}

\begin{thebibliography}{19}%
\makeatletter
\providecommand \@ifxundefined [1]{%
 \@ifx{#1\undefined}
}%
\providecommand \@ifnum [1]{%
 \ifnum #1\expandafter \@firstoftwo
 \else \expandafter \@secondoftwo
 \fi
}%
\providecommand \@ifx [1]{%
 \ifx #1\expandafter \@firstoftwo
 \else \expandafter \@secondoftwo
 \fi
}%
\providecommand \natexlab [1]{#1}%
\providecommand \enquote  [1]{``#1''}%
\providecommand \bibnamefont  [1]{#1}%
\providecommand \bibfnamefont [1]{#1}%
\providecommand \citenamefont [1]{#1}%
\providecommand \href@noop [0]{\@secondoftwo}%
\providecommand \href [0]{\begingroup \@sanitize@url \@href}%
\providecommand \@href[1]{\@@startlink{#1}\@@href}%
\providecommand \@@href[1]{\endgroup#1\@@endlink}%
\providecommand \@sanitize@url [0]{\catcode `\\12\catcode `\$12\catcode
  `\&12\catcode `\#12\catcode `\^12\catcode `\_12\catcode `\%12\relax}%
\providecommand \@@startlink[1]{}%
\providecommand \@@endlink[0]{}%
\providecommand \url  [0]{\begingroup\@sanitize@url \@url }%
\providecommand \@url [1]{\endgroup\@href {#1}{\urlprefix }}%
\providecommand \urlprefix  [0]{URL }%
\providecommand \Eprint [0]{\href }%
\providecommand \doibase [0]{http://dx.doi.org/}%
\providecommand \selectlanguage [0]{\@gobble}%
\providecommand \bibinfo  [0]{\@secondoftwo}%
\providecommand \bibfield  [0]{\@secondoftwo}%
\providecommand \translation [1]{[#1]}%
\providecommand \BibitemOpen [0]{}%
\providecommand \bibitemStop [0]{}%
\providecommand \bibitemNoStop [0]{.\EOS\space}%
\providecommand \EOS [0]{\spacefactor3000\relax}%
\providecommand \BibitemShut  [1]{\csname bibitem#1\endcsname}%
\let\auto@bib@innerbib\@empty
%</preamble>
\bibitem [{\citenamefont {Helstrom}(1976)}]{helstrom}%
  \BibitemOpen
  \bibfield  {author} {\bibinfo {author} {\bibfnamefont {C.~W.}\ \bibnamefont
  {Helstrom}},\ }\href {http://nla.gov.au/nla.cat-vn617918} {\emph {\bibinfo
  {title} {Quantum detection and estimation theory}}}\ (\bibinfo  {publisher}
  {Academic Press: New York},\ \bibinfo {year} {1976})\BibitemShut {NoStop}%
\bibitem [{\citenamefont {Bergou}\ \emph {et~al.}(2004)\citenamefont {Bergou},
  \citenamefont {Herzog},\ and\ \citenamefont {Hillery}}]{bergoubk}%
  \BibitemOpen
  \bibfield  {author} {\bibinfo {author} {\bibfnamefont {J.~A.}\ \bibnamefont
  {Bergou}}, \bibinfo {author} {\bibfnamefont {U.}~\bibnamefont {Herzog}}, \
  and\ \bibinfo {author} {\bibfnamefont {M.}~\bibnamefont {Hillery}},\
  }\href@noop {} {\emph {\bibinfo {title} {Quantum state estimation}}}\
  (\bibinfo  {publisher} {Springer, Berlin},\ \bibinfo {year} {2004})\ \bibinfo
  {type} {Chapter}~\bibinfo {chapter} {11}\BibitemShut {NoStop}%
\bibitem [{\citenamefont {Ivanovic}(1987)}]{Ivanovic1987257}%
  \BibitemOpen
  \bibfield  {author} {\bibinfo {author} {\bibfnamefont {I.~D.}\ \bibnamefont
  {Ivanovic}},\ }\href {\doibase 10.1016/0375-9601(87)90222-2} {\bibfield
  {journal} {\bibinfo  {journal} {Phys. Lett. A}\ }\textbf {\bibinfo {volume}
  {123}},\ \bibinfo {pages} {257 } (\bibinfo {year} {1987})}\BibitemShut
  {NoStop}%
\bibitem [{\citenamefont {Dieks}(1988)}]{Dieks1988303}%
  \BibitemOpen
  \bibfield  {author} {\bibinfo {author} {\bibfnamefont {D.}~\bibnamefont
  {Dieks}},\ }\href {\doibase 10.1016/0375-9601(88)90840-7} {\bibfield
  {journal} {\bibinfo  {journal} {Phys. Lett. A}\ }\textbf {\bibinfo {volume}
  {126}},\ \bibinfo {pages} {303 } (\bibinfo {year} {1988})}\BibitemShut
  {NoStop}%
\bibitem [{\citenamefont {Peres}(1988)}]{Peres198819}%
  \BibitemOpen
  \bibfield  {author} {\bibinfo {author} {\bibfnamefont {A.}~\bibnamefont
  {Peres}},\ }\href {\doibase 10.1016/0375-9601(88)91034-1} {\bibfield
  {journal} {\bibinfo  {journal} {Phys. Lett. A}\ }\textbf {\bibinfo {volume}
  {128}},\ \bibinfo {pages} {19 } (\bibinfo {year} {1988})}\BibitemShut
  {NoStop}%
\bibitem [{\citenamefont {van Enk}(2002)}]{PhysRevA.66.042313}%
  \BibitemOpen
  \bibfield  {author} {\bibinfo {author} {\bibfnamefont {S.~J.}\ \bibnamefont
  {van Enk}},\ }\href {\doibase 10.1103/PhysRevA.66.042313} {\bibfield
  {journal} {\bibinfo  {journal} {Phys. Rev. A}\ }\textbf {\bibinfo {volume}
  {66}},\ \bibinfo {pages} {042313} (\bibinfo {year} {2002})}\BibitemShut
  {NoStop}%
\bibitem [{\citenamefont {Bennett}(1992)}]{PhysRevLett.68.3121}%
  \BibitemOpen
  \bibfield  {author} {\bibinfo {author} {\bibfnamefont {C.~H.}\ \bibnamefont
  {Bennett}},\ }\href {\doibase 10.1103/PhysRevLett.68.3121} {\bibfield
  {journal} {\bibinfo  {journal} {Phys. Rev. Lett.}\ }\textbf {\bibinfo
  {volume} {68}},\ \bibinfo {pages} {3121} (\bibinfo {year}
  {1992})}\BibitemShut {NoStop}%
\bibitem [{\citenamefont {Delgado}\ \emph {et~al.}(2005)\citenamefont
  {Delgado}, \citenamefont {Roa}, \citenamefont {Retamal},\ and\ \citenamefont
  {Saavedra}}]{PhysRevA.71.012303}%
  \BibitemOpen
  \bibfield  {author} {\bibinfo {author} {\bibfnamefont {A.}~\bibnamefont
  {Delgado}}, \bibinfo {author} {\bibfnamefont {L.}~\bibnamefont {Roa}},
  \bibinfo {author} {\bibfnamefont {J.~C.}\ \bibnamefont {Retamal}}, \ and\
  \bibinfo {author} {\bibfnamefont {C.}~\bibnamefont {Saavedra}},\ }\href
  {\doibase 10.1103/PhysRevA.71.012303} {\bibfield  {journal} {\bibinfo
  {journal} {Phys. Rev. A}\ }\textbf {\bibinfo {volume} {71}},\ \bibinfo
  {pages} {012303} (\bibinfo {year} {2005})}\BibitemShut {NoStop}%
\bibitem [{\citenamefont {Clarke}\ \emph {et~al.}(2001)\citenamefont {Clarke},
  \citenamefont {Chefles}, \citenamefont {Barnett},\ and\ \citenamefont
  {Riis}}]{PhysRevA.63.040305}%
  \BibitemOpen
  \bibfield  {author} {\bibinfo {author} {\bibfnamefont {R.~B.~M.}\
  \bibnamefont {Clarke}}, \bibinfo {author} {\bibfnamefont {A.}~\bibnamefont
  {Chefles}}, \bibinfo {author} {\bibfnamefont {S.~M.}\ \bibnamefont
  {Barnett}}, \ and\ \bibinfo {author} {\bibfnamefont {E.}~\bibnamefont
  {Riis}},\ }\href {\doibase 10.1103/PhysRevA.63.040305} {\bibfield  {journal}
  {\bibinfo  {journal} {Phys. Rev. A}\ }\textbf {\bibinfo {volume} {63}},\
  \bibinfo {pages} {040305} (\bibinfo {year} {2001})}\BibitemShut {NoStop}%
\bibitem [{\citenamefont {Dada}\ \emph {et~al.}(2011)\citenamefont {Dada},
  \citenamefont {Andersson}, \citenamefont {Jones}, \citenamefont {Kendon},\
  and\ \citenamefont {Everitt}}]{PhysRevA.83.042339}%
  \BibitemOpen
  \bibfield  {author} {\bibinfo {author} {\bibfnamefont {A.~C.}\ \bibnamefont
  {Dada}}, \bibinfo {author} {\bibfnamefont {E.}~\bibnamefont {Andersson}},
  \bibinfo {author} {\bibfnamefont {M.~L.}\ \bibnamefont {Jones}}, \bibinfo
  {author} {\bibfnamefont {V.~M.}\ \bibnamefont {Kendon}}, \ and\ \bibinfo
  {author} {\bibfnamefont {M.~S.}\ \bibnamefont {Everitt}},\ }\href {\doibase
  10.1103/PhysRevA.83.042339} {\bibfield  {journal} {\bibinfo  {journal} {Phys.
  Rev. A}\ }\textbf {\bibinfo {volume} {83}},\ \bibinfo {pages} {042339}
  (\bibinfo {year} {2011})}\BibitemShut {NoStop}%
\bibitem [{\citenamefont {Neumann}\ \emph {et~al.}(2010)\citenamefont
  {Neumann}, \citenamefont {Beck}, \citenamefont {Steiner}, \citenamefont
  {Rempp}, \citenamefont {Fedder}, \citenamefont {Hemmer}, \citenamefont
  {Wrachtrup},\ and\ \citenamefont {Jelezko}}]{PhilippNeumann07302010}%
  \BibitemOpen
  \bibfield  {author} {\bibinfo {author} {\bibfnamefont {P.}~\bibnamefont
  {Neumann}}, \bibinfo {author} {\bibfnamefont {J.}~\bibnamefont {Beck}},
  \bibinfo {author} {\bibfnamefont {M.}~\bibnamefont {Steiner}}, \bibinfo
  {author} {\bibfnamefont {F.}~\bibnamefont {Rempp}}, \bibinfo {author}
  {\bibfnamefont {H.}~\bibnamefont {Fedder}}, \bibinfo {author} {\bibfnamefont
  {P.~R.}\ \bibnamefont {Hemmer}}, \bibinfo {author} {\bibfnamefont
  {J.}~\bibnamefont {Wrachtrup}}, \ and\ \bibinfo {author} {\bibfnamefont
  {F.}~\bibnamefont {Jelezko}},\ }\href {\doibase 10.1126/science.1189075}
  {\bibfield  {journal} {\bibinfo  {journal} {Science}\ }\textbf {\bibinfo
  {volume} {329}},\ \bibinfo {pages} {542} (\bibinfo {year}
  {2010})}\BibitemShut {NoStop}%
\bibitem [{\citenamefont {Gruber}\ \emph {et~al.}(1997)\citenamefont {Gruber},
  \citenamefont {Draebenstedt}, \citenamefont {Tietz}, \citenamefont {Fleury},
  \citenamefont {Wrachtrup},\ and\ \citenamefont
  {Borczyskowski}}]{Gruber_Science1997}%
  \BibitemOpen
  \bibfield  {author} {\bibinfo {author} {\bibfnamefont {A.}~\bibnamefont
  {Gruber}}, \bibinfo {author} {\bibfnamefont {A.}~\bibnamefont
  {Draebenstedt}}, \bibinfo {author} {\bibfnamefont {C.}~\bibnamefont {Tietz}},
  \bibinfo {author} {\bibfnamefont {L.}~\bibnamefont {Fleury}}, \bibinfo
  {author} {\bibfnamefont {J.}~\bibnamefont {Wrachtrup}}, \ and\ \bibinfo
  {author} {\bibfnamefont {C.~v.}\ \bibnamefont {Borczyskowski}},\ }\href
  {\doibase 10.1126/science.276.5321.2012} {\bibfield  {journal} {\bibinfo
  {journal} {Science}\ }\textbf {\bibinfo {volume} {276}},\ \bibinfo {pages}
  {2012} (\bibinfo {year} {1997})}\BibitemShut {NoStop}%
\bibitem [{\citenamefont {Jelezko}\ \emph {et~al.}(2004)\citenamefont
  {Jelezko}, \citenamefont {Gaebel}, \citenamefont {Popa}, \citenamefont
  {Domhan}, \citenamefont {Gruber},\ and\ \citenamefont
  {Wrachtrup}}]{Jelezko_PRL2004}%
  \BibitemOpen
  \bibfield  {author} {\bibinfo {author} {\bibfnamefont {F.}~\bibnamefont
  {Jelezko}}, \bibinfo {author} {\bibfnamefont {T.}~\bibnamefont {Gaebel}},
  \bibinfo {author} {\bibfnamefont {I.}~\bibnamefont {Popa}}, \bibinfo {author}
  {\bibfnamefont {M.}~\bibnamefont {Domhan}}, \bibinfo {author} {\bibfnamefont
  {A.}~\bibnamefont {Gruber}}, \ and\ \bibinfo {author} {\bibfnamefont
  {J.}~\bibnamefont {Wrachtrup}},\ }\href
  {http://link.aps.org/doi/10.1103/PhysRevLett.93.130501} {\bibfield  {journal}
  {\bibinfo  {journal} {Phys. Rev. Lett.}\ }\textbf {\bibinfo {volume} {93}},\
  \bibinfo {pages} {130501} (\bibinfo {year} {2004})}\BibitemShut {NoStop}%
\bibitem [{\citenamefont {Childress}\ \emph {et~al.}(2006)\citenamefont
  {Childress}, \citenamefont {Gurudev~Dutt}, \citenamefont {Taylor},
  \citenamefont {Zibrov}, \citenamefont {Jelezko}, \citenamefont {Wrachtrup},
  \citenamefont {Hemmer},\ and\ \citenamefont {Lukin}}]{Childress_Science2006}%
  \BibitemOpen
  \bibfield  {author} {\bibinfo {author} {\bibfnamefont {L.}~\bibnamefont
  {Childress}}, \bibinfo {author} {\bibfnamefont {M.~V.}\ \bibnamefont
  {Gurudev~Dutt}}, \bibinfo {author} {\bibfnamefont {J.~M.}\ \bibnamefont
  {Taylor}}, \bibinfo {author} {\bibfnamefont {A.~S.}\ \bibnamefont {Zibrov}},
  \bibinfo {author} {\bibfnamefont {F.}~\bibnamefont {Jelezko}}, \bibinfo
  {author} {\bibfnamefont {J.}~\bibnamefont {Wrachtrup}}, \bibinfo {author}
  {\bibfnamefont {P.~R.}\ \bibnamefont {Hemmer}}, \ and\ \bibinfo {author}
  {\bibfnamefont {M.~D.}\ \bibnamefont {Lukin}},\ }\href {\doibase
  10.1126/science.1131871} {\bibfield  {journal} {\bibinfo  {journal}
  {Science}\ }\textbf {\bibinfo {volume} {314}},\ \bibinfo {pages} {281}
  (\bibinfo {year} {2006})}\BibitemShut {NoStop}%
\bibitem [{\citenamefont {Balasubramanian}\ \emph {et~al.}(2009)\citenamefont
  {Balasubramanian}, \citenamefont {Neumann}, \citenamefont {Twitchen},
  \citenamefont {Markham}, \citenamefont {Kolesov}, \citenamefont {Mizuochi},
  \citenamefont {Isoya}, \citenamefont {Achard}, \citenamefont {Beck},
  \citenamefont {Tissler}, \citenamefont {Jacques}, \citenamefont {Hemmer},
  \citenamefont {Jelezko},\ and\ \citenamefont {Wrachtrup}}]{Balasubramanian}%
  \BibitemOpen
  \bibfield  {author} {\bibinfo {author} {\bibfnamefont {G.}~\bibnamefont
  {Balasubramanian}}, \bibinfo {author} {\bibfnamefont {P.}~\bibnamefont
  {Neumann}}, \bibinfo {author} {\bibfnamefont {D.}~\bibnamefont {Twitchen}},
  \bibinfo {author} {\bibfnamefont {M.}~\bibnamefont {Markham}}, \bibinfo
  {author} {\bibfnamefont {R.}~\bibnamefont {Kolesov}}, \bibinfo {author}
  {\bibfnamefont {N.}~\bibnamefont {Mizuochi}}, \bibinfo {author}
  {\bibfnamefont {J.}~\bibnamefont {Isoya}}, \bibinfo {author} {\bibfnamefont
  {J.}~\bibnamefont {Achard}}, \bibinfo {author} {\bibfnamefont
  {J.}~\bibnamefont {Beck}}, \bibinfo {author} {\bibfnamefont {J.}~\bibnamefont
  {Tissler}}, \bibinfo {author} {\bibfnamefont {V.}~\bibnamefont {Jacques}},
  \bibinfo {author} {\bibfnamefont {P.~R.}\ \bibnamefont {Hemmer}}, \bibinfo
  {author} {\bibfnamefont {F.}~\bibnamefont {Jelezko}}, \ and\ \bibinfo
  {author} {\bibfnamefont {J.}~\bibnamefont {Wrachtrup}},\ }\href {\doibase
  10.1038/nmat2420} {\bibfield  {journal} {\bibinfo  {journal} {Nature
  Materials}\ }\textbf {\bibinfo {volume} {8}},\ \bibinfo {pages} {383}
  (\bibinfo {year} {2009})}\BibitemShut {NoStop}%
\bibitem [{\citenamefont {Robledo}\ \emph {et~al.}(2011)\citenamefont
  {Robledo}, \citenamefont {Childress}, \citenamefont {Bernien}, \citenamefont
  {Hensen}, \citenamefont {Alkemade},\ and\ \citenamefont
  {Hanson}}]{Robledo_Nature2011}%
  \BibitemOpen
  \bibfield  {author} {\bibinfo {author} {\bibfnamefont {L.}~\bibnamefont
  {Robledo}}, \bibinfo {author} {\bibfnamefont {L.}~\bibnamefont {Childress}},
  \bibinfo {author} {\bibfnamefont {H.}~\bibnamefont {Bernien}}, \bibinfo
  {author} {\bibfnamefont {B.}~\bibnamefont {Hensen}}, \bibinfo {author}
  {\bibfnamefont {P.~F.~A.}\ \bibnamefont {Alkemade}}, \ and\ \bibinfo {author}
  {\bibfnamefont {R.}~\bibnamefont {Hanson}},\ }\href
  {http://dx.doi.org/10.1038/nature10401} {\bibfield  {journal} {\bibinfo
  {journal} {Nature}\ }\textbf {\bibinfo {volume} {477}},\ \bibinfo {pages}
  {574} (\bibinfo {year} {2011})}\BibitemShut {NoStop}%
\bibitem [{\citenamefont {Waldherr}\ \emph {et~al.}(2011)\citenamefont
  {Waldherr}, \citenamefont {Neumann}, \citenamefont {Huelga}, \citenamefont
  {Jelezko},\ and\ \citenamefont {Wrachtrup}}]{Waldherr_PRL2011b}%
  \BibitemOpen
  \bibfield  {author} {\bibinfo {author} {\bibfnamefont {G.}~\bibnamefont
  {Waldherr}}, \bibinfo {author} {\bibfnamefont {P.}~\bibnamefont {Neumann}},
  \bibinfo {author} {\bibfnamefont {S.~F.}\ \bibnamefont {Huelga}}, \bibinfo
  {author} {\bibfnamefont {F.}~\bibnamefont {Jelezko}}, \ and\ \bibinfo
  {author} {\bibfnamefont {J.}~\bibnamefont {Wrachtrup}},\ }\href {\doibase
  10.1103/PhysRevLett.107.090401} {\bibfield  {journal} {\bibinfo  {journal}
  {Phys. Rev. Lett.}\ }\textbf {\bibinfo {volume} {107}},\ \bibinfo {pages}
  {090401} (\bibinfo {year} {2011})}\BibitemShut {NoStop}%
\bibitem [{\citenamefont {Franke-Arnold}\ \emph {et~al.}(2001)\citenamefont
  {Franke-Arnold}, \citenamefont {Andersson}, \citenamefont {Barnett},\ and\
  \citenamefont {Stenholm}}]{PhysRevA.63.052301}%
  \BibitemOpen
  \bibfield  {author} {\bibinfo {author} {\bibfnamefont {S.}~\bibnamefont
  {Franke-Arnold}}, \bibinfo {author} {\bibfnamefont {E.}~\bibnamefont
  {Andersson}}, \bibinfo {author} {\bibfnamefont {S.~M.}\ \bibnamefont
  {Barnett}}, \ and\ \bibinfo {author} {\bibfnamefont {S.}~\bibnamefont
  {Stenholm}},\ }\href {\doibase 10.1103/PhysRevA.63.052301} {\bibfield
  {journal} {\bibinfo  {journal} {Phys. Rev. A}\ }\textbf {\bibinfo {volume}
  {63}},\ \bibinfo {pages} {052301} (\bibinfo {year} {2001})}\BibitemShut
  {NoStop}%
\bibitem [{\citenamefont {Reck}\ \emph {et~al.}(1994)\citenamefont {Reck},
  \citenamefont {Zeilinger}, \citenamefont {Bernstein},\ and\ \citenamefont
  {Bertani}}]{PhysRevLett.73.58}%
  \BibitemOpen
  \bibfield  {author} {\bibinfo {author} {\bibfnamefont {M.}~\bibnamefont
  {Reck}}, \bibinfo {author} {\bibfnamefont {A.}~\bibnamefont {Zeilinger}},
  \bibinfo {author} {\bibfnamefont {H.~J.}\ \bibnamefont {Bernstein}}, \ and\
  \bibinfo {author} {\bibfnamefont {P.}~\bibnamefont {Bertani}},\ }\href
  {\doibase 10.1103/PhysRevLett.73.58} {\bibfield  {journal} {\bibinfo
  {journal} {Phys. Rev. Lett.}\ }\textbf {\bibinfo {volume} {73}},\ \bibinfo
  {pages} {58} (\bibinfo {year} {1994})}\BibitemShut {NoStop}%
\end{thebibliography}
%==================================================================================================================
%

%===========================================================================================
\end{document}